\documentclass[10pt,conference]{IEEEtran}
\usepackage{hyperref}
\usepackage{mdframed}
\usepackage{lipsum}
\usepackage{amsmath}
\usepackage{amssymb}
\usepackage{amsfonts}
\usepackage{ifpdf}
\usepackage{todonotes}
\usepackage{booktabs} 
\usepackage{url}
\usepackage{amssymb}
\usepackage{pifont}
\usepackage{tcolorbox}
\usepackage{balance}
\usepackage{makecell}
\usepackage{verbatim}
\usepackage{rotating}

\usepackage{bussproofs}
\usepackage{lipsum} 
\usepackage{mathpartir}
\usepackage{algorithmic}
\usepackage{listings}
\usepackage{array}

\usepackage{caption}
\usepackage{multirow}
\usepackage{float}
\usepackage{mwe}
\newcommand{\code}[1]{{\fontfamily{cmtt}\fontseries{m}\fontshape{n}\selectfont\small{#1}}}

\usepackage{url}
\usepackage{listings}
\usepackage{todonotes}
\usepackage{subfigure}

\usepackage{tikz}
\usepackage{xcolor}
\definecolor{dkgreen}{rgb}{0,0.6,0}

\usepackage[para]{threeparttable}

\makeatletter
\newcommand{\linebreakand}{%
  \end{@IEEEauthorhalign}
  \hfill\mbox{}\par
  \mbox{}\hfill\begin{@IEEEauthorhalign}
}
\makeatother

\usepackage{algorithm2e}
\usepackage{xcolor}

\SetAlFnt{\footnotesize}

\SetAlCapSty{xAlCapSty}

\SetCommentSty{xCommentSty}

\SetNlSty{mynlfont}{}{} 

\LinesNumbered

\SetSideCommentRight

\DontPrintSemicolon

\RestyleAlgo{algoruled}

\begin{document}

\title{Towards Black-box Attacks on Deep Learning Apps}

\author{\IEEEauthorblockN{Hongchen Cao}
\IEEEauthorblockA{
ShanghaiTech University\\
caohch1@shanghaitech.edu.cn}
\and
\IEEEauthorblockN{Shuai Li}
\IEEEauthorblockA{
The Hong Kong Polytechnic University\\
csshuaili@comp.polyu.edu.hk}
\and
\IEEEauthorblockN{Yuming Zhou}
\IEEEauthorblockA{
Nanjing University\\
zhouyuming@nju.edu.cn}
\linebreakand 
\IEEEauthorblockN{Ming Fan}
\IEEEauthorblockA{
Xi’an Jiaotong University\\
mingfan@xjtu.edu.cn}
\and
\IEEEauthorblockN{Xuejiao Zhao}
\IEEEauthorblockA{
Nanyang Technological University\\
xjzhao@ntu.edu.sg}
\and
\IEEEauthorblockN{Yutian Tang*}
\IEEEauthorblockA{
ShanghaiTech University\\
tangyt1@shanghaitech.edu.cn}
}

\maketitle

\begin{abstract}

Deep learning is a powerful weapon to boost application performance in various fields, including face recognition, object detection, image classification, natural language understanding, and recommendation system. With the rapid increase in the computing power of mobile devices, developers can embed deep learning models into their apps for building more competitive products with more accurate and faster responses. Although there are several works of adversarial attacks against deep learning models in apps, they all need information about the models' internals (i.e., structures, weights) or need to modify the models. In this paper, we propose an effective black-box approach by training substitute models to spoof the deep learning systems inside the apps. We evaluate our approach on 10 real-world deep-learning apps from Google Play to perform black-box adversarial attacks. Through the study, we find three factors that can affect the performance of attacks. Our approach can reach a relatively high attack success rate of 66.60\% on average. Compared with other adversarial attacks on mobile deep learning models, in terms of the average attack success rates, our approach outperforms counterparts by 27.63\%.

\end{abstract}

\begin{IEEEkeywords}
Black-box attacks,
Deep Learning apps,
Android
\end{IEEEkeywords}

\IEEEpeerreviewmaketitle

\section{Introduction}
The explosive progress of deep learning techniques makes an increasing number of app developers pay attention to this technology. Deep learning apps reach a certain scale in the market and keep growing rapidly~\cite{xu2019first}. To lower the threshold for developers to utilize and deploy deep learning models on their apps, companies (i.e., Google) develop deep learning frameworks, including Tensorflow Lite~\cite{tensorflowlite}, Pytorch Mobile~\cite{pytorchmobile}, Caffe2 Mobile~\cite{caff2mobile}, MindSpore Lite~\cite{mindsporelite}, Core ML~\cite{coreml} and so on.


To build a deep learning app, mainstream platforms offer two common deploying strategies, on-device deployment and on-cloud deployment~\cite{firebaseOndeviceVSOncloud}. On-cloud deployment allows developers to deploy their models on the remote server, and collect the runtime inputs from app users. Then, the inputs are processed by the remote server. Finally, the results are returned to app users. However, the efficiency of this method is often limited by the network quality and power usage~\cite{mcintosh2019can}. Even worse, on-cloud deployment has potential risks of user privacy leakage~\cite{kumar2020adversary}. Therefore, lots of developers turn to the on-device deployment strategy.

\noindent\textbf{Motivation.} The aforementioned mainstream mobile deep learning frameworks and deployment platforms rarely provide developers with solutions or APIs to protect their deep learning models. Unprotected deep learning models are widely used in apps in various fields such as finance, health, and self-driving~\cite{Sun2020MindYW}. This may lead to potential user data leakage, commercial information theft, and the malfunction of important functional modules of the app (e.g. a polluted face recoginition model fails to recoginize an authorized user).

\noindent\textbf{State-of-art.} Although some studies~\cite{li2021deeppayload} focus on attacks on mobile deep learning models, their approaches are often white-box, which means that the attacker can know or even modify the structure (the number, type, and arrangement of layers) and weights (parameters of each layer) of the model. These white-box attacks rely on the source of victim models, which can be extracted by decompressing the victim app. However, a recent study shows that developers are more likely to use encryption or hash code to protect their models~\cite{Sun2020MindYW}. It means that the knowledge of a model (i.e., structure and weights) inside the app is always unknown, which nullifies these white-box attacks. Work~\cite{huang2021robustness} claims to be a black-box attack method towards deep learning models. However, they still require knowledge of the structure or weight of the model. Such information is commonly unknown in the real world, especially for commercial apps.

\noindent\textbf{Our solution.} In this paper, we developed a practical pipeline that covers data preparation, student model (i.e., a substitute model to mimic the behavior of the victim model) learning, and adversarial examples generation. Our black-box attack approach can bypass many existing model protection measures. First, we instrument a \code{DummyActivity} to the victim app to invoke the deep learning model (a.k.a teacher model) adopted by the victim app. The raw data fetched from the Internet or public dataset are fed into the teacher model through the \code{DummyActivity} to generate the corresponding labels (See Sec. \ref{subsec:dataset-prepartion}). Next, a substitute model (a.k.a. student model), which is selected to mimic the behavior of the teacher model, is then trained with the labeled data (See Sec. \ref{subsec:model-learning}). Last, adversarial attack algorithms are applied to the student model, and the generated adversarial examples are leveraged to attack the teacher model in the victim app (See Sec. \ref{subsec:adversarial-generation}).

\noindent\textbf{Contributions.} In summary, our main contributions are described as follows:


\noindent$\bullet$ We propose a black-box attack pipeline, from the model collection to adversarial example generation, to attack deep learning apps. 

\noindent$\bullet$ We conduct a series of experiments to evaluate how different factors affect the attack success rate, including the structure of the student models, the scale of the teaching dataset, and the adversarial attack algorithms; and 

\noindent$\bullet$ We propose a series of strategies to defend against the proposed attacks.

We release the source code and data on the online artefact~\cite{onlineartefacts}.

\section{Background}
\subsection{On-device Deep Learning Model}
Developers are increasingly using deep learning models to implement face recognition~\cite{yin1029feature}, image classification~\cite{daniel2010towards}, object detection~\cite{daniel2010towards}, natural language understanding~\cite{chen2020unblind}, speech recognition~\cite{dong2020rtmobile}, and recommendation systems~\cite{wang2020next} in their apps. 

To meet the needs of app developers, vendors build deep learning frameworks for mobile platforms, such as TensorFlow Lite~\cite{tensorflowlite} and PyTorch Mobile~\cite{pytorchmobile}. They also provide platforms (e.g. TensorFlow Hub~\cite{tensorflowhub}, MindSpore Model~\cite{mindspore}) to share pre-trained deep learning models for mobile apps. 


There are two common deploying strategies, on-device deployment and on-cloud deployment. Developers nowadays are inclined to use the on-device method to deploy deep learning models. On the one hand, the hardware of modern mobile devices is qualified to complete computationally intensive deep learning tasks. As the CPU and GPU of modern mobile devices have stronger computing power, on-device inference can be completed promptly. The neural processing units (NPU) in mobile devices further expand the feasibility of on-device deployment of deep learning models~\cite{tan2020fastva}.  

On the other hand, deep learning frameworks provide simple and easy-to-use APIs to invoke the model for inference tasks. For example, as shown in List. \ref{list:tfinterpreter}, \textit{Interpreter} defined in the TensorFlow Lite is used to load the model and run inference. \textit{file\_of\_a\_tflite\_model} represents the file path of the deep learning model. \textit{input} presents the user's input and \textit{output} represents the result of the inference which is a vector. Based on the output, the app can give a response to the user. For example, the user feeds an image of a rose as \textit{input} into the \textit{Interpreter}. After inference, the \textit{output} is assigned to a one-hot vector like [0, 0, 1, 0, ..., 0]. According to pre-determined rules from developers, the vector is then interpreted into a human-readable format such as a string "rose".

\begin{lstlisting}[label={list:tfinterpreter}, caption=TensorFlow Lite API]
# TensorFlow Lite
Interpreter interpreter = new Interpreter(file_of_a_tflite_model);
interpreter.run(input, output);
\end{lstlisting}

In the apps, developers can define the locations of their deep learning models. A common practice is to store the models in the assets folder (/assets/) of the apk (Android Package, which is the binary format of an app). Data including user input, output vector, and training dataset are invisible to users. Although models in some deep learning apps can be directly extracted by decompiling the apk, lots of them still cannot be fetched even with the state-of-art tools~\cite{Sun2020MindYW}. Therefore, the information about the structures and weights of these deep learning models is unknown to attackers.

\subsection{Black-box Attack}
Since the information about the victim is always unknown, attacks must be performed in a black-box manner. A black-box attack is not a specific attack technique but describes an attack scenario. In this scenario, the attackers have no internal knowledge of the victim, and cannot modify the internal structure and other attributes of the victim.
In our context, the attackers have no internal knowledge of the deep learning model in the victim app and cannot modify the internal attributes of the model. The attackers can only query the model and get its prediction.


\subsection{Adversarial Attack}

Deep learning models are vulnerable to adversarial attacks. The adversarial attack utilizes adversarial examples to spoof the deep learning models. An adversarial example is generated by adding tiny perturbations to the input data. It can be represented as $x_{adv} = x + \epsilon * p$, where $x_{adv}$ is the adversarial example, $x$ is the input data, $p$ is the perturbations and $\epsilon$ is the multiplier to control the degree of the perturbations~\cite{goodfellow2015explaining, huang2021robustness}. An adversarial attack can be briefly summarized in the following example. With a deep learning model for image classification $M(x)$ and an input image $x$,  $y_{true}$ is the label of $x$ ($y_{true}=M(x)$). Attackers can generate an adversarial example $x'$ by adding infinitesimal perturbations that a human being cannot distinguish from the original image $x$. But the adversarial example $x'$ can spoof the victim deep learning model to make an incorrect prediction ($M(x') \neq y_{true}$). 

\begin{figure}[!htbp]
    \centering
    \subfigure[orignal input]{
    \includegraphics[width=0.21\textwidth]{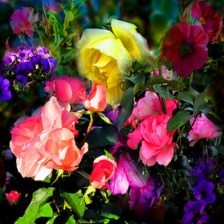}
    \label{img:background-1}
    }
    \quad
    \subfigure[adversarial example]{
    \includegraphics[width=0.21\textwidth]{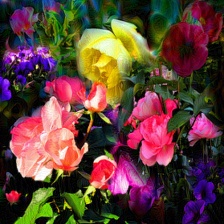}
    \label{img:background-2}
    }
    \vspace{-0.5em}
    \caption{Comparsion for orignal input and adversarial example}
    \vspace{-0.5em}
    \label{img:backgroung-example}
\end{figure}

\begin{figure*}[!htpb]
	\centering
	\includegraphics[trim=0 100 0 100,width=\textwidth]{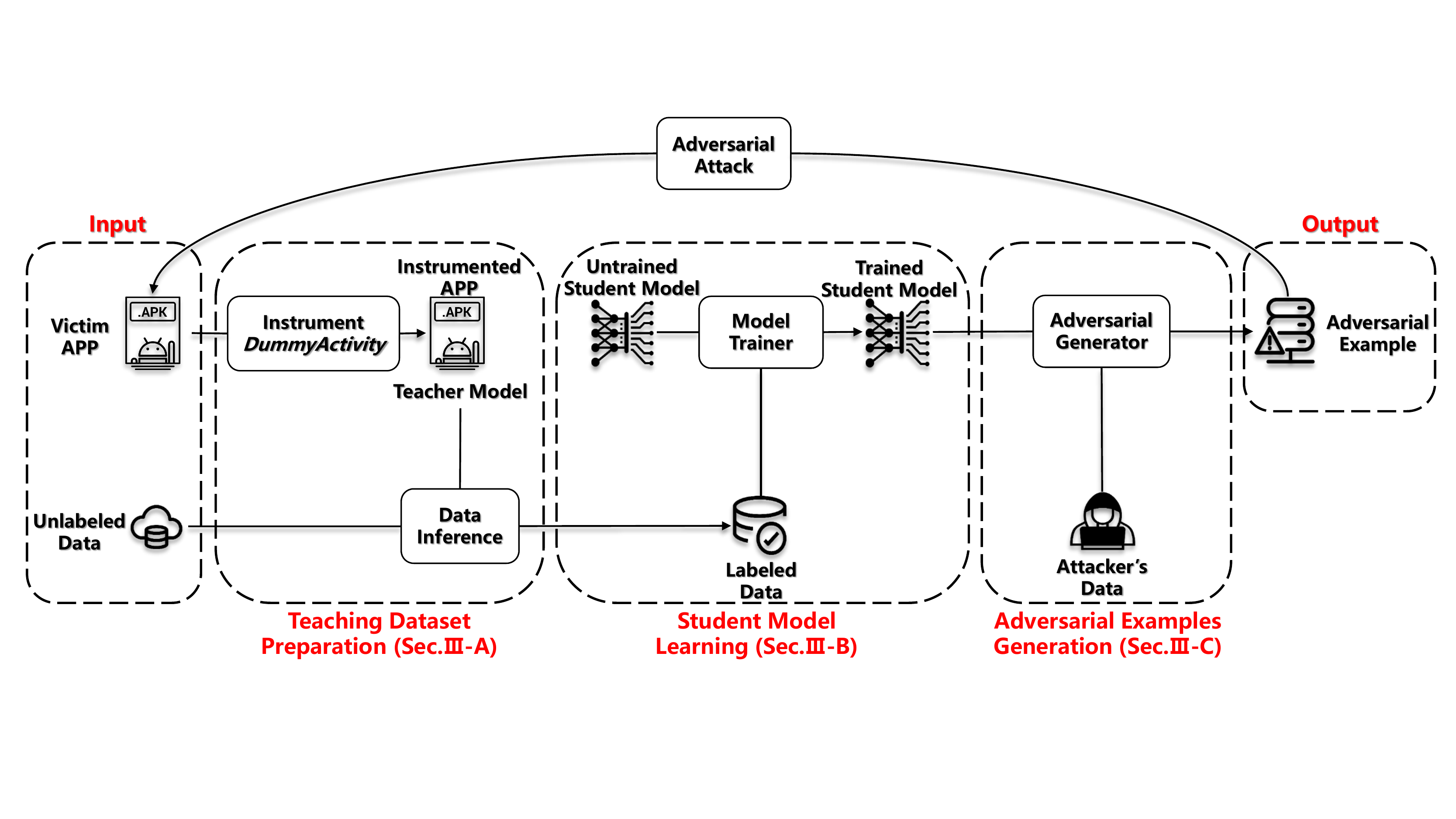}
	\vspace{-0.5em}
	\caption{Overview of adversarial attack pipeline}
	\vspace{-0.5em}
	\label{img:pipeline}
\end{figure*}

For example, Fig. \ref{img:background-1} is the original input image, which is labeled as 'rose' by a deep learning model $M$. After applying adversarial attacks to it, an adversarial example is generated as shown in Fig. \ref{img:background-2}, which is labeled as "tulips" by $M$.

\section{Methodology}\label{sec:methodlogy}

Before introducing the details of our approach, we first define two key concepts used in our work: teacher model and student model. A teacher model is the deep learning model that is used in the victim app. A student model can be considered as a substitute or imitator of the teacher model. A qualified student model can imitate the teacher’s behaviors well. With the same input, the student and the teacher model are supposed to output the same prediction. A well-behaved student model can guarantee that the adversarial examples generated can be directly transferred to the teacher model.

Our black-box attack pipeline, as illustrated in Fig. \ref{img:pipeline}, consists of three procedures: teaching dataset preparation, student model learning, and adversarial examples generation.

\noindent$\bullet$ \textbf{Input:} The input of our pipeline is a deep learning app and an unlabeled dataset. The dataset is either crawled from the web or a public dataset that is related to the app's task. For example, the dataset for an app whose task is flower classification can be images of different kinds of flowers.

\noindent$\bullet$ \textbf{Teaching dataset preparation:} Given a deep learning app, we first locate and learn how the deep learning model is invoked by tracking the official APIs from deep learning frameworks in the app. Then, we instrument a \code{DummyActivity}, in which the teacher model is invoked, to the victim app with Soot \cite{Soot,flowdroid}. Next, we feed the unlabeled dataset into the \code{DummyActivity} to generate the corresponding labels (see Sec. \ref{subsec:dataset-prepartion}).

\noindent$\bullet$ \textbf{Student model learning:} To imitate the teacher model, we select mainstream deep learning models with the same task as the student model candidates. With the labeled teaching dataset generated, a student model can be trained to imitate the teacher model  (see Sec. \ref{subsec:model-learning}).

\noindent$\bullet$ \textbf{Adversarial examples generation:} With the trained student model, adversarial examples can be generated with adversarial attack algorithms (see Sec. \ref{subsec:adversarial-generation}).

\noindent$\bullet$ \textbf{Output:} The output of the whole pipeline is a set of adversarial examples, which are used to attack the victim app.

\subsection{Teaching Dataset Preparation}\label{subsec:dataset-prepartion}
For a deep learning app, we identify some basic attributes of the teacher model in the app. These attributes mainly consist of the type of its task, the name of classes it predicts, and the formatting requirement of the input. 

\noindent$\bullet$\textbf{Type of task and Name of classes }: Determining the task of the deep learning model is a prerequisite for collecting a suitable dataset. The name of the output (i.e., classes) of the deep learning model can further help narrow the selection range of the dataset. This information can be obtained by trying the app.

\noindent$\bullet$\textbf{Formatting requirement of the input }: The input format of the deep learning model is determined by the model developer. These formatting requirements include size (e.g., 128*128), data type (e.g., int8, float32), and so forth. By reverse engineering the app, these requirements can be obtained manually.

For a plant identifier app, its task is image classification. Its names of classes can be the names of plants (e.g, lily, plane tree). Its formatting requirement of the input can be a 255$\times$255$\times$3 RGB image.

Based on this information, we search a public dataset $D$ from the web. However, it is common that no public related dataset is available for real-world apps. There are two options to collect the dataset for such apps. The first is to use a pre-trained student model without fine-tuning by the teaching dataset, which is simple and fast. The second is to crawl the related data from the Internet. We compare these two options with experiments in Sec. \ref{subsec:rq2-dataset}.

Since the input format of the deep learning model is determined by the model developer, the collected dataset needs to be preprocessed to make it meet the specific formatting requirements. Based on the previously obtained information about the input formatting requirements, the input data can be manipulated (e.g., zoom in size, update figure accuracy) to meet the input formatting requirements.

The dataset we crawled is unlabeled. However, to better imitate the teacher model, we need to train the student model with a labeled dataset. Therefore, we need to extract the labels of the dataset with the teacher model. For a black-box attack, it can be unrealistic to directly retrieve the teacher model from the app. Therefore, we perform the following steps to transfer an unlabeled dataset $D$ into a labeled one $D'$:

\noindent$\bullet$ STEP 1: We collect the API patterns of mainstream mobile deep learning model deployment frameworks. The corresponding APIs can be found in the official docs (i.e., Tensorflow Lite inference~\cite{tensorflowliteinference}, Pytorch Mobile inference~\cite{pytorchinference}).

\noindent$\bullet$ STEP 2: We use Soot to locate the signatures of these APIs in the app. 

\noindent$\bullet$ STEP 3: We instrument \code{DummyActivity} into the app with Soot and load the deep learning model inside the \code{onCreate()} method of \code{DummyActivity} with the same API found in STEP 2.

\noindent$\bullet$ STEP 4: We pass the dataset $D$ to the inference API and output the corresponding labels with logs.

\noindent$\bullet$ STEP 5: We leverage the Android Debug Bridge (ADB) to launch our \code{DummyActivity} (i.e., \code{am start -n DummyActivity}) to trigger the teacher model. As a result, we can obtain the corresponding labels for the inputs.

\noindent\textbf{Example.} The Mushroom Identifier app (\code{com.gabotechindustries.mushroomIdentifier}) detects and identifies mushrooms in the picture and predicts their categories. This app has a class called "Xception". List. \ref{list:mushroomexample} summarizes how this app invokes the deep learning model in "Xception" with the APIs provided by TensorFlow Lite.

\begin{lstlisting}[float, label=list:mushroomexample, basicstyle=\small,caption=API pattern of invoking the deep learning model]
public class Xception {
    public Xception(..., final String str, ...) {
        ...
        final Interpreter.Options options = new Interpreter.Options();
        this.tflite = new Interpreter(map, options);}
}
public abstract class AIActivity extends CameraActivity {
    protected void onCreate(...) {
        this.xception = new Xception(..., "xception103.tflite", ...);}
}
\end{lstlisting}

As illustrated in List. \ref{list:dummyactivity}, we embed the extracted pattern (in List. \ref{list:mushroomexample}) into \code{DummyActivity}. Then, we instrument the \code{DummayActivity} to the app.

\begin{lstlisting}[float, label=list:dummyactivity, basicstyle=\small,caption=Instrumented DummyActivity]
class DummyActivity extends AppCompatActivity {
 public void onCreate(Bundle savedInstanceState) {
 ...
 // Init and run the Interpreter
 Interpreter interpreter = new Interpreter("xception103.tflite");
 interpreter.run(input, output); 
 int maxIdx = output.indexOf(output.max());//Get label
 Log.i("DummyActivity",maxIdx.toString());//Output label
 }
 ...
}
\end{lstlisting}

After instrumenting the \code{DummyActivity} to the app, we launch the \code{DummyActivity} with an ADB command (i.e., \code{am start -n DummyActivity}). As a result, the corresponding labels of $D$ can be obtained. The labeled dataset $D'$ is then leveraged to train the student model.

\subsection{Student Model Learning}\label{subsec:model-learning}
\noindent\textbf{Selection criteria for student model.}
The adversarial examples generated based on the student model are used to attack the teacher model. The selection criteria of the student model is that it should be able to complete the same task as the teacher model. Different student models can result in different attack success rates. We discuss this correlation in Sec. \ref{subsec:rq1-studentstructure}.


\noindent\textbf{Student model learning.} In our approach, the student model can be trained in a supervised way with the teaching dataset generated in the previous step. In supervised learning, the model is trained on a labeled dataset, where the labels are used to evaluate the model's prediction accuracy. To improve the prediction accuracy, the model needs to adjust the weights in the process of training. The adjusting strategy relies on the error between the model's predictions and the ground truths(i.e., labels). The loss function is used to calculate this error. For example, the CrossEntropyLoss is widely used in image classification~\cite{DBLP:conf/gcce/TakedaYM20}.



\noindent\textbf{Example.} Recall the app in Sec. \ref{subsec:dataset-prepartion}, its task is to identify mushrooms in an input image and predict their categories. Thus, the selected student model must be a model related to image classification. The image classification model VGGll~\cite{simonyan2014very} can be used as the student model to be trained with the labeled teaching dataset generated in Sec. \ref{subsec:dataset-prepartion}. Note that a general image classification model can be trained to identify mushrooms with a suitable dataset.

\subsection{Adversarial Examples Generation}\label{subsec:adversarial-generation}

With the trained student model in the previous step, the next step is to use the student model to generate adversarial examples that can spoof the teacher model. The adversarial examples are proved to have the property of transferability~\cite{liu2017delving} which means that the adversarial examples generated for one model through an attack algorithm can also attack another model. Motivated by this, adversarial attack algorithms are utilized to generate adversarial examples based on the student model, and these generated adversarial examples can be directly applied to attack the teacher model in the victim app. Note that different domains (e.g., NLP, computer vision) require different adversarial attack algorithms. For example, in computer vision, the following algorithms can be representative: FGSM~\cite{goodfellow2015explaining}, RFGSM~\cite{tramer2020ensemble}, FFGSM~\cite{kim2020torchattacks}, MIFGSM~\cite{dong2018boosting}, PGD~\cite{madry2017towards}, TPGD~\cite{zhang2019theoretically}, and BIM~\cite{kurakin2016adversarial}. Different attack algorithms can result in different attack performances. We discuss this correlation in Sec. \ref{subsec:rq3-hyperpara}.

\begin{figure*}[!htbp]
	\centering
	\includegraphics[trim=0 0 40 45, width=0.9\textwidth]{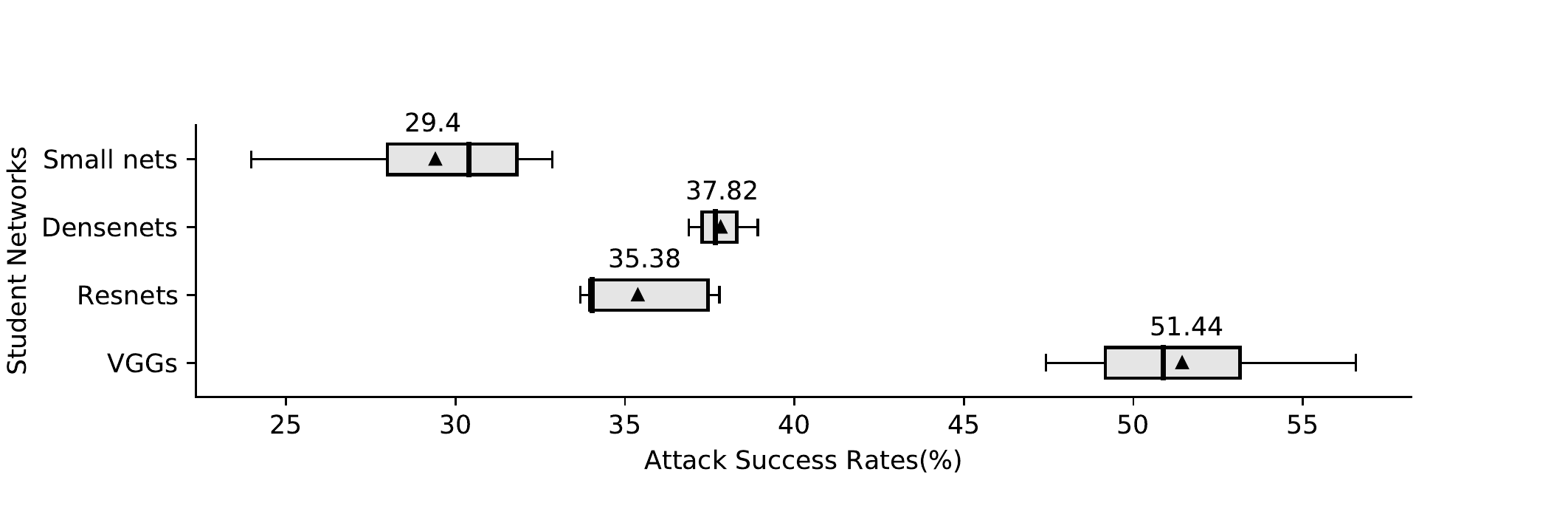}
	\vspace{-0.6em}
	\caption{Distributions of attack success rate for differnet models}
	\vspace{-0.6em}
	\label{img:rq1-box}
\end{figure*}



\subsection{Robustness Evaluation}\label{subsec: robustness-evaluation}
Given a teacher model, the teaching dataset is divided into training and testing sets. The training set is used to train the student model and the testing set is used to generate adversarial examples. Compared with current methods~\cite{huang2021robustness} that use a very limited number of attack examples, our evaluation set is much larger and thus our results are more convincing.

Suppose there are total $M$ inputs in the testing set for a deep learning app. $T$ (out of $M$) inputs can be correctly classified by the teacher model. These $T$ inputs are used to generate adversarial examples since it is inappropriate to generate an adversarial example when the input cannot be correctly recognized by the teacher model. Then we generate $T$ adversarial examples with the student model through an existing attack algorithm (e.g., FGSM~\cite{goodfellow2015explaining}). These adversarial examples are then fed to the teacher model to obtain the prediction results. An attack is considered to be successful if the adversarial example is misclassified by the teacher model.
If $Q$ out of $T$ adversarial examples are misclassified by the teacher model, then the attack success rate is:
\begin{equation}
p=\frac{Q}{T}
\end{equation}

The quality of the student model is key to reach a high attack success rate. A qualified student model can imitate the behavior of the teacher model well and generate adversarial examples with higher transferability. These adversarial examples are more likely to spoof the teacher model~\cite{liu2017delving}. Therefore, we conduct experiments to investigate the influence factors on the attack success rate of the student model, including the scale of the teaching dataset and the model structure of the student model. To further evaluate the effectiveness of our method, we conduct one control experiment that uses a random pre-trained model to generate adversarial examples termed as ‘blind attack’ (see Sec. \ref{subsec:rq4-finalresulets}).





\section{Evaluation}

\subsection{Experiment Settings}

In the following experiments, We reuse 10 deep learning apps in the work of Huang et al.~\cite{huang2021robustness} so that we can compare our approach with theirs. Considering the workload of training student models and applying different attack methods to generate adversarial examples, it is not trivial to experiment on 10 apps. For example, one teaching dataset is ImageNet Large Scale Visual Recognition Challenge 2012 (ILSVRC2012)~\cite{imagenet} whose size is above 138GB and the number of images is about 14 million. This takes more than a week for a computer with 8 NVIDIA Tesla K40s.

\begin{table}[!htbp]
\centering
\caption{10 selected apps}
\scalebox{1}{
\begin{tabular}{|c|c|c|}
\hline
\textbf{ID} & \textbf{App Name}            & \textbf{Functionalities}            \\ \hline
1           & Fresh Fruits                          & Identify if the fruit is rotten    \\ \hline
2           & Image Checker                         & Image classifier based on ImageNet \\ \hline
3           & Tencent Map                           & Identify road condition            \\ \hline
4           & Baidu Image                           & Identify flowers                    \\ \hline
5           & My Pokemon                            & Identify pokemon                   \\ \hline
6           & Palm Doctor                           & Identify skin cancer               \\ \hline
7           & QQ browser                                 & Identify plants                               \\ \hline
8           & Taobao                                  & Identify products                               \\ \hline
9           & iQIYI                                  & Identify actors             \\ \hline
10          & Bei Ke                                  & Identify scenes           \\ \hline
\end{tabular}
}
\label{table:collected-model}
\end{table}

The selected apps are listed in Table.\ref{table:collected-model}. For convenience, in the following experiments, every app is represented by its ID instead of name. Deep learning functionalities of each app are also listed, which are useful for finding suitable dataset. 

Research questions (RQ) 1 to 3 discuss the relationship between attack performance and student structure (RQ1), teaching dataset size (RQ2), and hyper-parameter of attack algorithm (RQ3). We compare the attack success rates between our approach with ~\cite{huang2021robustness} in RQ4. All these experiments\footnote{In the process of training student models, we set optimizer to SGD~\cite{torchsgd}, loss function to CrossEntropyLoss~\cite{torchcross}, learning rate to 0.001, and epoch to 30.} are conducted in a Linux box with Nvidia(R) GeForce RTX(TM) 3080 GPU and 128G RAM.

\subsection{RQ1:  How the structure of the student model influences the attack rate.}\label{subsec:rq1-studentstructure}

\begin{figure*}[!htpb]
    \centering
    \subfigure[Relationship of attack success rate and teaching dataset size]{
    \includegraphics[trim=0 0 5 5, width=0.45\textwidth]{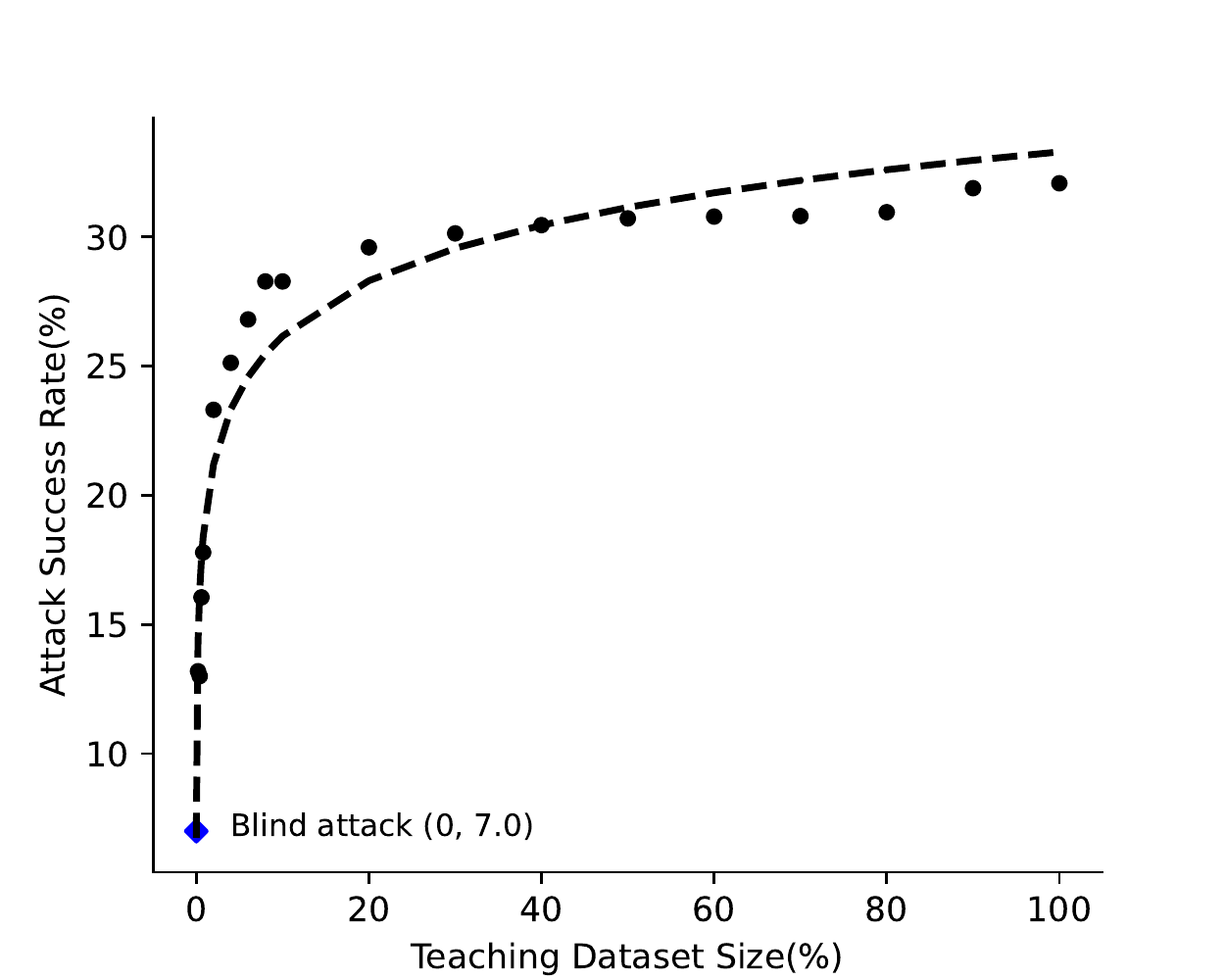}
    \label{img:rq2-scatter1}
    }
    \quad
    \subfigure[Relationship of student model accuracy and teaching dataset size]{
    \includegraphics[trim=0 0 5 5, width=0.45\textwidth]{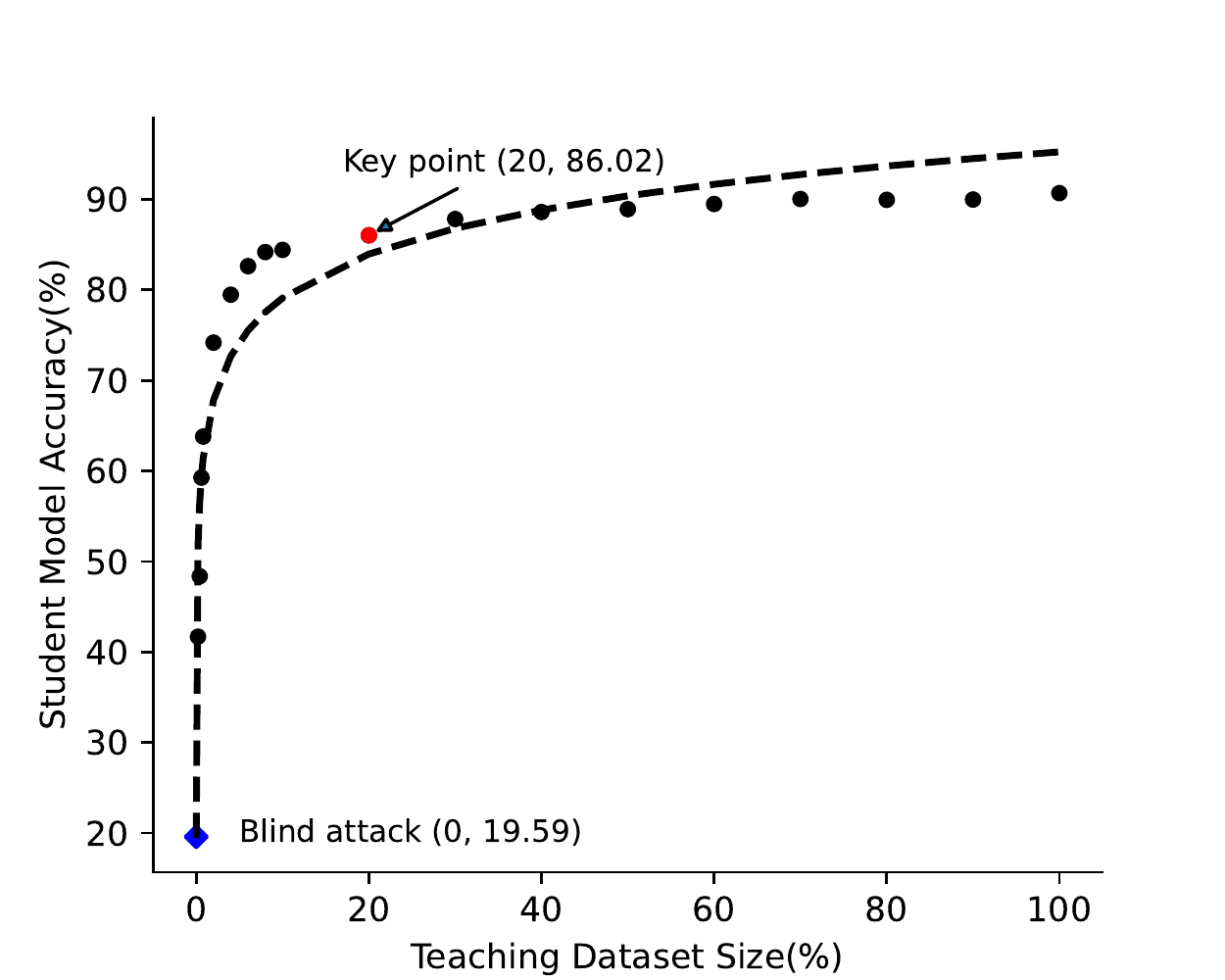}
     \label{img:rq2-scatter2}
    }
    \caption{Performance of attack and student model when varying teaching dataset size}
    \label{img:rq2-scatter}
\end{figure*}
 
\noindent\textbf{Motivation.} The stronger the student model's ability to imitate the teacher model is, the higher the final attack success rate is. Since the internal information of the teacher model is unknown, choosing the most similar student model by comparing weights and structure can be impossible. Therefore, in this RQ, we intend to explore how the structure of the student model influences the attack rate and analyze the underlying reasons. The attack rate is defined in Sec. \ref{subsec: robustness-evaluation}.

\noindent\textbf{Approach.} This RQ explores how the structure of the student model influences the attack rate, so only one teacher model and only one attack algorithm are needed. In this RQ, we randomly select one app (i.e., No. 1). The deep learning model inside the app is considered as the teacher model, MIFGSM as the attack algorithm. The teaching dataset is a dataset from Kaggle \cite{kaggledatasetfruit} with 13.6K images. The teaching dataset is shuffled and then divided into training and testing sets at 4:1. We carefully select 16 student models that are qualified for this image classification task. These student models are grouped into four categories with the decrease of their model sizes and complexity:

\begin{itemize}
    \item VGGs~\cite{simonyan2014very}: VGG11, VGG13, VGG16, VGG19;
    \item Resnets~\cite{kai2016deep}: Resnet18, Resnet34, Resnet50, Resnet101, Resnet152;
    \item Densenets~\cite{huang2017densely}: Densenet161, Densenet169, Densenet201; and
    \item Small nets: MobilenetV2~\cite{mark2018mobilenet}, ShufflenetV2~\cite{ma2018shufflenet}, Squeezenet~\cite{iandola2016squeezenet}, Googlenet~\cite{szegedy2015going}.
\end{itemize}

Among the above four categories, VGGs have the highest complexity in terms of the number of weights, which is 3-4 times that of Resnets and Densenets, and 20-30 times that of Small nets. Resnets and Densenets have similar sizes and complexity.

\noindent\textbf{Result.} Fig. \ref{img:rq1-box} shows the comparison of attack performance among four student model categories. The abscissa represents attack success rates in percentage, and the ordinate gives four categories. In every single box, the triangle represents the mean of attack success rate and the line represents the median. The leftmost and rightmost edges represent minimum and maximum attack success rates in the corresponding categories. 

The attack success rates are divided into three segments. The first watershed is between Small nets and Resnets, the average attack success rate of Small nets is 29.4\%, the minimum is 23.97\% ($Q:T=637:2657$, where $Q$ and $T$ are defined in Sec. \ref{subsec: robustness-evaluation}), and the maximum is 32.86\% ($873:2657$). The second watershed is between Densenets and VGGS. Densenets and Resnets who have similar model complexities also have similar attack success rates. Densenets' average attack success rate is 37.82\%, the minimum is 36.88\% ($980:2657$), and the maximum is 38.92\% ($1034:2657$). Resnets' average attack success rate is 35.28\%, the minimum is 33.68\% ($895:2657$), and the maximum is 37.79\% ($1004:2657$). The most successful student models are VGGs. Their average attack success rate is 51.44\%, the minimum is 47.42\% ($1260:2657$), and the maximum is 56.57\% ($1503:2657$).

Throughout the experiment, we found that a more sophisticated student model has a better opportunity and flexibility to adjust its weights to better imitate the teacher model. The better the student model imitates, the stronger the transferability of the adversarial examples generated by the student model is.
 
\noindent\textbf{Num. of apps.} Due to page restriction, we only present the result of one app. We also perform the same experiments on other apps and observe the same trend on them.

\begin{tcolorbox}[title=Answer to RQ1,boxrule=1pt,boxsep=1pt,left=2pt,right=2pt,top=2pt,bottom=2pt]
To better imitate the teacher model and reach a higher attack success rate, a student model with a more complex structure is a better choice. Among 16 models in 4 categories, VGGs have a relatively high average attack success rate compared with other student model candidates.
\end{tcolorbox} 

\subsection{RQ2: How the scale of the teaching dataset can influence the attack rate. }\label{subsec:rq2-dataset}

\begin{figure*}[!htbp]
	\centering
	\includegraphics[trim=10 0 10 5,width=\textwidth]{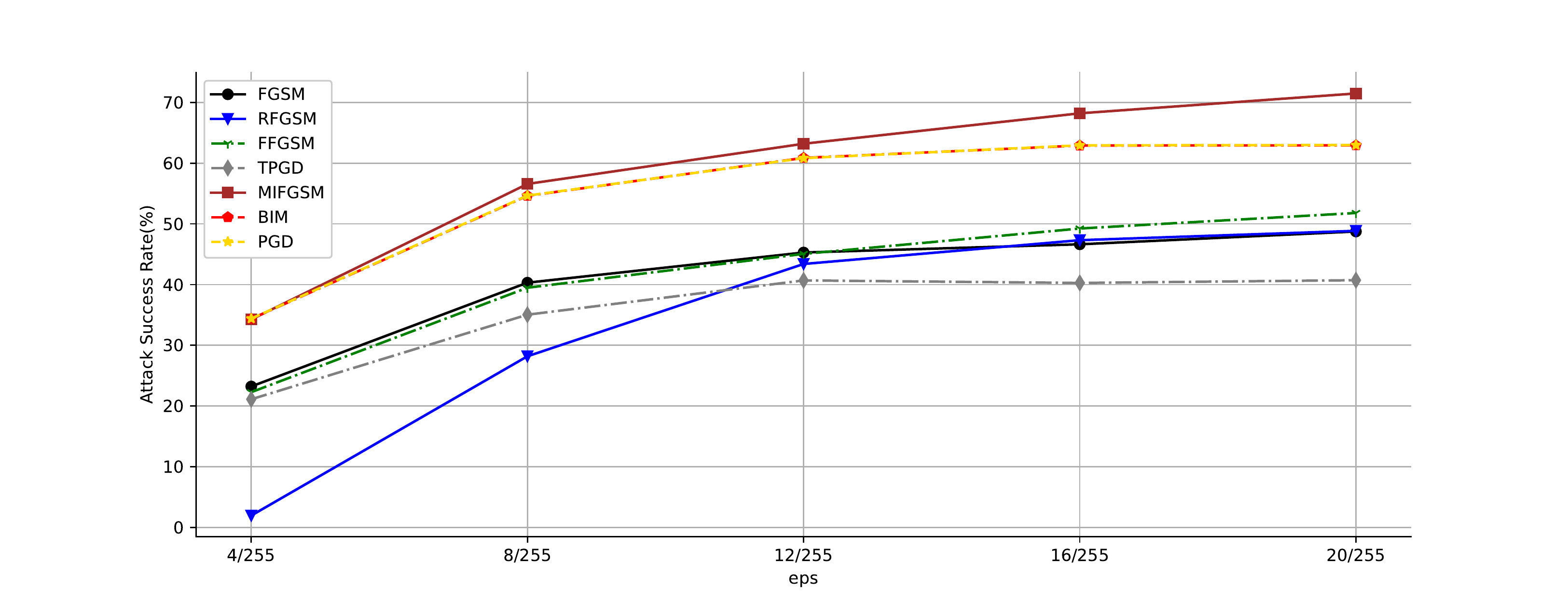}
	\vspace{-1em}
	\caption{Relationship between attack success rate and paramater \textit{eps}}
	\vspace{-1em}
	\label{img:rq3-line}
\end{figure*}

\noindent\textbf{Motivation. } 
Training a student model on a large teaching dataset consumes lots of time and computing resources. The size of the teaching dataset can influence the prediction accuracy of the student model. The prediction accuracy reflects how well the student model imitates the teacher model. Thus, this RQ explores how the scale of the teaching dataset can influence the attack rate.

\noindent\textbf{Approach. }
As this RQ explores how the scale of the teaching dataset influences the attack success rate, so only one teacher model, one attack algorithm, and one structure of the student model are needed. In this RQ, we randomly select one app (i.e., No.4) as the teacher model, MIFGSM as the attack algorithm, and VGG11 trained on a teaching dataset from TensorFlow~\cite{tfdataset} with 3.6K images as the student model.

The teaching dataset is shuffled and then divided into training and testing sets at 4:1. The attack success rate and student model accuracy of each student model is evaluated on the same testing set. training sets are obtained from the left part of the teaching dataset. Note that without training, the method degenerated to ‘blind attack'.

\noindent\textbf{Result.}
Fig. \ref{img:rq2-scatter1} shows the relationship between the size of the teaching dataset (i.e., abscissa in Fig. \ref{img:rq2-scatter1}) and attack success rate (i.e., ordinate in Fig. \ref{img:rq2-scatter1}). For example, point (10, 28.28) represents that the attack success rate of the student model trained on the teaching dataset with 367 images (10\% of original size) is 28.28\%.

Compared with the models trained on our teaching dataset (black points), the performance of blind attack (blue points) is 2-4 times worse. It proves that a teaching dataset generated by our pipeline is indispensable. To better reflect the relationship, trend line is based on a logarithmic function $y=a \cdot ln(b + x) + c$, where $a$, $b$ and $c$ control how well the trend line fits the scattered points. The trend line shows that the attack success rate increases rapidly and gradually stabilizes with the teaching dataset increases. With the increase of the scale of the teaching dataset, the student model can further adjust its weights, resulting in better imitation of the teacher model and higher transferability of generated adversarial examples.

The reason for the gradual stabilization of the trend line needs to be explained with the help of Fig. \ref{img:rq2-scatter2}. The abscissa in Fig. \ref{img:rq2-scatter2} represents the teaching dataset size, which is the percentage of the original dataset, and the ordinate represents the accuracy of the student model in percentage. For example, point (20, 86.02) represents that the prediction accuracy of the student model trained on the teaching dataset with 687 images (20\% of original size) is 86.02\%. Compared with other cases, a model without training (i.e., (0, 19.59) in Fig. \ref{img:rq2-scatter2}) has a poor performance. (20, 86.02) is a key point, after which the accuracy of the student model becomes stable. 

Higher accuracy means a higher model similarity between the student model and the teacher model. The key point (20, 86.02) in Fig. \ref{img:rq2-scatter2} represents that the imitation ability of the student model is close to its upper bound. Therefore, further increasing the teaching dataset brings negligible improvement in terms of the attack success rate. This is why the growth of attack success rate in Fig. \ref{img:rq2-scatter1} becomes stable after this key point. 

\noindent\textbf{Num. of apps.} Same as RQ1, we perform the same experiments on other apps and observe the same trend on other apps.


\begin{tcolorbox}[title=Answer to RQ2,boxrule=1pt,boxsep=1pt,left=2pt,right=2pt,top=2pt,bottom=2pt]
A teaching dataset generated by our pipeline is necessary for a high attack success rate. The accuracy of the student model increases as the size of the teaching dataset grows. The accuracy of the student model reflects how well it imitates the teacher model. When the accuracy of the student model reaches around 85\%-90\%, its imitation ability is close to the upper bound. Therefore, blindly increasing the teaching dataset contribute less to the attack performance.
\end{tcolorbox}

\subsection{RQ3: How the hyper-parameter of attack algorithms influences the attack performance.}\label{subsec:rq3-hyperpara}

\begin{figure*}[!htbp]
    
    \centering
    \subfigure[MIFGSM with $eps=4/255$]{
    \includegraphics[scale=0.35]{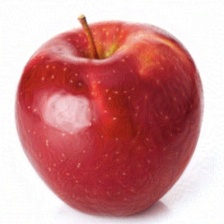}
    }
    \quad
    \subfigure[MIFGSM with $eps=8/255$]{
    \includegraphics[scale=0.35]{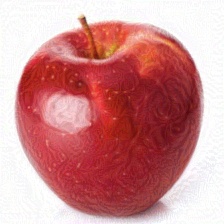}
    }
    \quad
    \subfigure[MIFGSM with $eps=12/255$]{
    \includegraphics[scale=0.35]{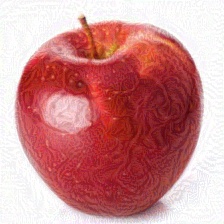}
    }
    \quad
    \subfigure[MIFGSM with $eps=16/255$]{
    \includegraphics[scale=0.35]{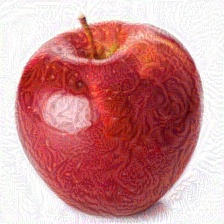}
    }
    \quad
    \subfigure[MIFGSM with $eps=20/255$]{
    \includegraphics[scale=0.35]{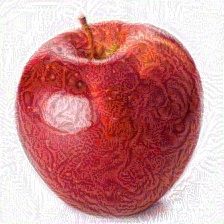}
    }
    \quad
    
    \subfigure[BIM with $eps=4/255$]{
    \includegraphics[scale=0.35]{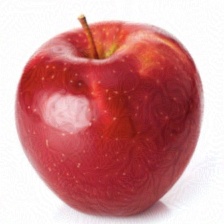}
    }
    \quad
    \subfigure[BIM with $eps=8/255$]{
    \includegraphics[scale=0.35]{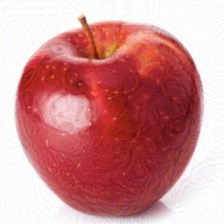}
    }
    \quad
    \subfigure[BIM with $eps=12/255$]{
    \includegraphics[scale=0.35]{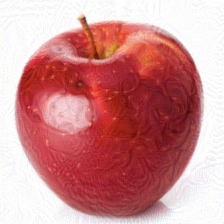}
    }
    \quad
    \subfigure[BIM with $eps=16/255$]{
    \includegraphics[scale=0.35]{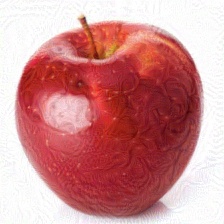}
    }
    \quad
    \subfigure[BIM with $eps=20/255$]{
    \includegraphics[scale=0.35]{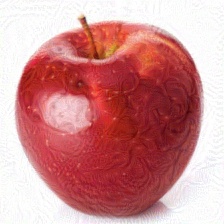}
    }
    \quad
    
    \subfigure[PGD with $eps=4/255$]{
    \includegraphics[scale=0.35]{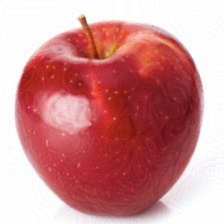}
    }
    \quad
    \subfigure[PGD with $eps=8/255$]{
    \includegraphics[scale=0.35]{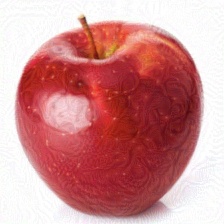}
    }
    \quad
    \subfigure[PGD with $eps=12/255$]{
    \includegraphics[scale=0.35]{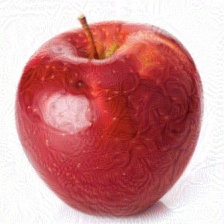}
    }
    \quad
    \subfigure[PGD with $eps=16/255$]{
    \includegraphics[scale=0.35]{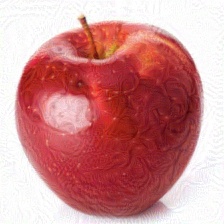}
    }
    \quad
    \subfigure[PGD with $eps=20/255$]{
    \includegraphics[scale=0.35]{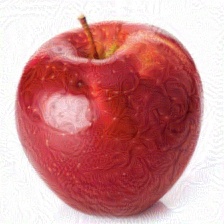}
    }
    \quad
    \vspace{-0.5em}
    \caption{Comparison between adversarial example generated with different eps}
    \vspace{-0.5em}
    \label{img:rq3-examples}
\end{figure*}

\noindent\textbf{Motivation. } The attack performance is evaluated on both the attack success rate and the degree of the added perturbation to the original input. A key property of the adversarial example is that it is generated by adding tiny perturbations to the original input. If perturbations are too large, they can be noticed by human beings. If the perturbations are too small, the attack success rate reduces. The following experiment explores how the hyper-parameters of attack algorithms can influence attack performance.

\noindent\textbf{Approach.} This RQ explores how the hyper-parameters of attack algorithms influences the attack performance, so only one teacher model and one student model are needed. In this RQ, we randomly select one app (i.e., No.1) as the teacher model, VGG11 trained on a teaching dataset from Kaggle~\cite{kaggledatasetfruit} with 13.6K images as the student model. 

The important hyper-parameter \textit{eps} controls the degree of image perturbation in all seven attack algorithms (FGSM, BIM, RFGSM, PGD, FFGSM, TPGD, MIFGSM). \textit{eps} varies from $4/255$ to $20/255$ with step $4/255$. The initial value $eps=4/255$ is a default value used by the author of these 7 attack algorithms~\cite{dong2018boosting, goodfellow2015explaining, tramer2020ensemble, kim2020torchattacks, madry2017towards, zhang2019theoretically, kurakin2016adversarial}. We set the end value to $eps=20/255$. Although larger $eps$ can bring a higher attack success rate, the perturbations of the image also become larger. A higher degree of perturbations brings a more significant difference between the original input image and the adversarial example. As a result, people can distinguish the difference between the original input and the adversarial example. To ensure the degree of the perturbations is minuscule, the maximum of \textit{eps} is set to $20/255$. This reduces the upper bound of the success rate but can ensure perturbations to be unperceivable. 

\noindent\textbf{Result.}
As shown in Fig. \ref{img:rq3-line}, the attack success rate of FGSM ranges from 23.22\% ($617:2657$) to 48.74\% ($1295:2657$), RFGSM ranges from 1.96\% ($52:2657$) to 48.85\% ($1298:2657$), FFGSM ranges from 22.28\% ($592:2657$) to 51.79\% ($1376:2657$), TPGD ranges from 21.11\% ($561:2657$) to 40.72\% ($1082:2657$), MIFGSM ranges from 34.25\% ($910:2657$) to 71.47\% ($1899:2657$), BIM ranges from 34.36\% ($913:2657$) to 62.93\% ($1672:2657$), and PGD ($914:2657$) ranges from 34.40\% ($1668:2657$) to 62.97\%. All algorithms reach its highest attack succes rate with $eps=20/255$. 

Fig. \ref{img:rq3-line} also indicates that different attack algorithms have different sensitivity to the adjustment of \textit{eps}. However,  $eps=8/255$ is an important watershed for all seven attack algorithms. When $eps \leq 8/255$, the attack success rate decrease rapidly. This is because that the perturbations are too minuscule to spoof the teacher model.

Another important watershed is $eps=16/255$. As shown in Fig. \ref{img:rq3-examples}, when $eps\geq16/255$ ($4^{th}$ and $5^{th}$ colomun in Fig. \ref{img:rq3-examples}) the adversarial examples have relatively large perturbations. Such perturbation can be detected by human beings, which causes the adversarial example unqualified. 

Since attack performance is determined by the success rate and the degree of perturbation, we evaluate the performance on both sides.

\noindent$\bullet$ \textbf{Attack success rate.} MIFGSM shows the highest attack success rate through all 7 attack algorithms, reaching 71.47\% when $eps=20/255$. Also, it can maintain a relatively rapid growth rate after the first watershed $eps=8/255$. The reason for its strong attack capability is that the adversarial examples generated by it are more transferable. Compared with other FGSM based attack algorithms, the highest attack success rates of FFGSM, RFGSM, and FGSM are 51.79\%, 48.85\%, and 48.74\%, respectively. The reason why MIFGSM outperforms other algorithms is that it invites the momentum term into the iterative process~\cite{dong2018boosting}.

\noindent$\bullet$ \textbf{Degree of perturbation.} Although the attack success rate of MIFGSM keeps increasing quickly after both watersheds, the perturbations in adversarial examples become detectable by the human visual system when $eps\geq16/255$. This behavior is common among all 4 FGSM based attack algorithms, so adjusting \textit{eps} to $16/255$ or higher is not recommended when using this type of algorithm. For BIM and PGD, the highest attack success rates of them are the same (62.93\%). Although the attack success rate is less than MIFGSM's, the degree of perturbation in their generated adversarial example is much smaller than MIFGSM. Therefore BIM and PGD with $eps\geq16/255$ are recommended.

\noindent\textbf{Num. of apps.} Same as RQ1, we perform the same experiments on other apps and observe the same trend on other apps.

\begin{tcolorbox}[title=Answer to RQ3,boxrule=1pt,boxsep=1pt,left=2pt,right=2pt,top=2pt,bottom=2pt]
An ideal \textit{eps} interval is [8/255, 16/255]. Keeping \textit{eps} in this interval can guarantee a higher attack success rate with limited added perturbation.
\end{tcolorbox} 

\subsection{RQ4: Comparison among our approach, ModelAttacker and blind attack}\label{subsec:rq4-finalresulets}

\noindent\textbf{Motivation.}
We compare the performance among our approach, \textit{ModelAttacker} proposed by Huang et al.~\cite{huang2021robustness} and blind attack on 10 apps in Table. \ref{table:collected-model}. The experiment settings of \textit{ModelAttacker} and blind attack are as same as our approach.

\noindent\textbf{Approach. }
The attack success rates of all apps in Table. \ref{table:collected-model} are tested with the best student model structure, the most cost-effective teaching dataset size, and attack algorithms with a suitable value of \textit{eps} (defined in Sec. \ref{subsec:rq3-hyperpara}).

\noindent\textbf{Teaching dataset.} The teaching datasets for most apps are fetched from public datasets, including Kaggle-Fruit~\cite{kaggledatasetfruit} (No. 1), ILSVRC2012~\cite{imagenet} (No. 2, 8, 10), Kaggle-Road~\cite{kaggledatasetroad} (No. 3), Tendorflow-Flowers~\cite{tfdataset} (No. 4), Kaggle-Pokemon~\cite{kaggledatasetpokemon} (No. 5), and HAM10000~\cite{hamdataset} (No. 6). For the other 2 apps (No. 7, 9), the teaching datasets are crawled from Google Images~\cite{googleimages}. These datasets are available in our online artefact~\cite{onlineartefacts}. Each teaching dataset is shuffled and then divided into training and testing sets at 4:1.

\noindent\textbf{Student model.} Based on the result of RQ1, we select VGG11 as the student model for all 10 apps.

\noindent\textbf{Attack algorithm.} Based on the result of RQ3, we select MIFGSM as the attack algorithm and set $eps$ to $12/255$. 

\begin{figure}[!htpb]
	\centering
	\includegraphics[trim=10 0 10 5,width=0.5\textwidth]{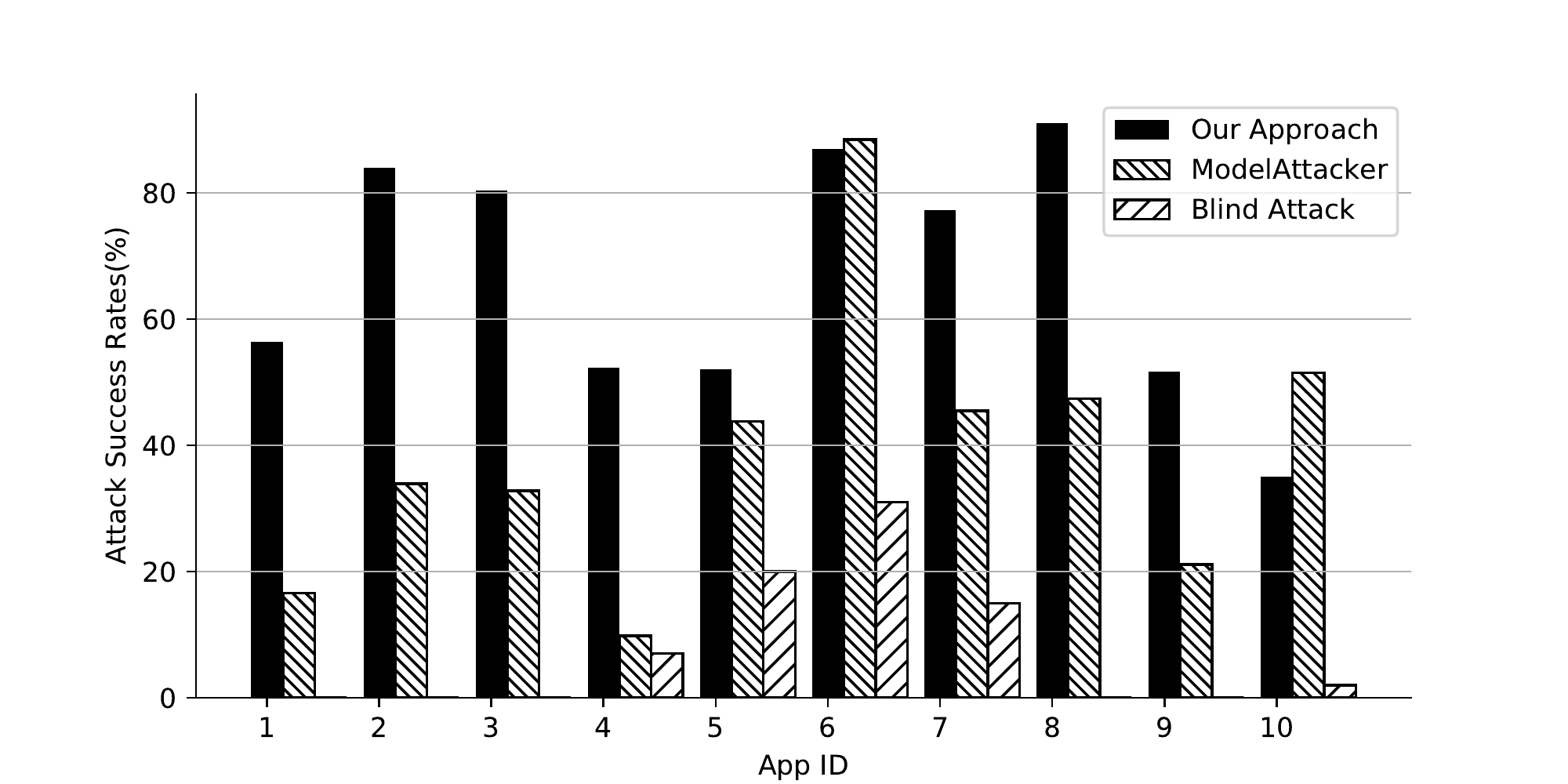}
	\caption{Comparison among our approach, \textit{ModelAttacker} and blind attack}
	\label{img:final-results}
\end{figure}

\noindent\textbf{Result.}
As shown in Fig. \ref{img:final-results}, the height of the first bar for each app represents our attack success rate, the second is for \textit{ModelAttacker}, and the third is for the blind attack. Apps with No.4,5,6,7,10 can be attacked by the blind attack so the other 5 apps do not have the bars to show the results of blind attack. 

The range of our attack success rate is from 34.98\% to 91.10\% and on average is 66.60\%. Our approach does not need the weights and structures of deep learning models inside the victim apps. Although with these constraints, our approach still gets a higher attack success rate for 8 of 10 apps compared with \textit{ModelAttacker}. Specifically, the attack success rate of our approach is 8.27\% to 50.03\% higher than \textit{ModelAttacker}, and on average is 36.79\% higher than \textit{ModelAttacker}. 

For the left 2 apps, \textit{ModelAttacker} can reach a higher attack success rate. This is because that they can compare the similarity between the victim model (i.e., teacher model) and their substitute model (i.e., student model) with the knowledge of weights and structure of the teacher model~\cite{huang2021robustness}. However, our approach is a black box attack so the weights and structure of the teacher model are unknown. 

Compared with blind attack which is the common basic black-box attack method, our approach can attack all 10 apps successfully while the blind attack can only attack 5 of them. Meanwhile, the attack success rate of our approach is at least 2.6 times higher than the blind attack.

\begin{tcolorbox}[title=Answer to RQ4,boxrule=1pt,boxsep=1pt,left=2pt,right=2pt,top=2pt,bottom=2pt]
Our approach can indeed effectively attack 10 selected deep learning apps, thus providing a new perspective and method for evaluating app reliability. Compared with existing methods, we can reach a relatively high attack success rate of 66.60\% on average, outperforming others by 27.63\%.
\end{tcolorbox} 

\section{Discussion}



\subsection{Threats to Validity}\label{limitation}
To minimize the bias of the results, all three experiments in this paper vary a single variable while fixing other variables to evaluate its impact on attack performance. Different from other studies on deep learning apps~\cite{li2021deeppayload,huang2021robustness}, our experiments are based on a larger dataset and generate a considerable number of adversarial examples for the victim app in each experiment.  Considering the workload of training, we select 10 apps that are also used by~\cite{huang2021robustness} to evaluate our approaches. Since the success rates of attacks are based on these apps, we do not attempt to generalize our results to all deep learning apps.

\subsection{Limitations}
Our study has two limitations. The main limitation is that our approach only suitable for non-obfuscated apps. To get the teaching dataset from a teacher model inside the app, we have to locate and learn how the model is loaded and used in the app. However, if protection techniques, such as code obfuscation, are applied to the app, it can be hard to find the API patterns. Another limitation is that the pipeline developed in this paper focuses on computer vision apps. On the one hand, the existing studies~\cite{huang2021robustness, li2021deeppayload} on adversarial attacks are mainly in this field. It can be easy to compare our work with other approaches. On the other hand, adversarial attacks in other deep learning fields lack a widely accepted evaluation standard. Without widely adopted and convincing criteria, it can be unfair to compare the results of our approach with others~\cite{dong2021sentence,wang2019natural}. Thus, we only consider computer vision apps in our experiments.

\subsection{Consequences of Proposed Attacks}
From the perspective of a malicious attacker, the adversarial examples can threaten users’ property, privacy, and even  safety. Dong et al.~\cite{dong2019efficient} applied adversarial examples to a real-world face recognition system successfully. As a popular way to unlock devices, the breach of the face recognition system means users' privacy is at risk of leakage. Although it is not easy to use adversarial examples in the real world, relevant research has made progress~\cite{sun2018survey, Boloor2019simple, Xu2020what}. The threat of adversarial examples is worthy of attention. 

\subsection{Countermeasure}

According to the study by Sun et al.~\cite{Sun2020MindYW}, only 59\% of deep learning apps have protection methods for their models. To fill this gap, we propose several protection methods for both deep learning models and deep learning apps. 

\noindent\textbf{Deep learning model protection scheme.} To protect the deep learning model inside an app from black-box attacks, there are two practical solutions:

\noindent$\bullet$ Using self-developed deep learning models instead of open-source models can reduce the probability of being attacked. For a self-developed deep learning model, it can be hard to find or train a qualified student model to imitate the teacher model. Without a qualified student model, it can be hard to reach a high attack success rate; and

\noindent$\bullet$ Training the deep learning model on a private dataset instead of a public dataset can reduce the success rate of being attacked. The size of the teaching dataset is critical for training a qualified student model (See Sec. \ref{subsec:rq2-dataset}). If the deep learning model inside the app is trained on a public dataset, the attacker can find the same dataset with ease. A student model can gain a powerful imitating ability by being trained on the same dataset as the teacher model. As a result, the adversarial examples generated on the student model can spoof the teacher model with a high success rate.

\noindent\textbf{Deep learning app protection scheme.} Using a self-developed deep learning model and collecting a private dataset is time-consuming and costly. There are also some common techniques to protect deep learning apps from being attacked~\cite{Sun2020MindYW, li2021deeppayload, huang2021robustness}:

\noindent$\bullet$ Code obfuscation can prevent attackers from finding out how the apps invoke the deep learning models. As a result, attackers cannot generate a labeled teaching dataset by instrumenting a \code{DummyActivity}. Thus, the attack method degenerates to the blind attack with a low attack success rate;

\noindent$\bullet$ Hash code can prevent attackers from modifying the model inside the victim app. Li et al.~\cite{li2021deeppayload} use images with special patterns to fool the models by adding specific layers into deep learning models inside apps. With hash code, such backdoor attacks can be detected with ease; and

\noindent$\bullet$ Encryption can prevent attackers from knowing the structure and weights of the model. It is a fundamental method to protect the deep learning model inside the app. Without the knowledge of the victim model's structure and weights, attackers can only rely on black-box attack methods, which can be time-consuming (i.e., finding dataset, training the student model).

\section{Related Work}

\subsection{Substitute-based adversarial attacks}
To overcome black-box attacks' unreachability to the internals of victim deep learning models, attackers can train a local model to mimic the victim model, which is called substitute training. With the in-depth study of black-box attacks on deep learning models, substitute-based adversarial attacks have received a lot of attention. Cui et al.~\cite{cui2020substitute} proposed an algorithm to generate the substitute model of CNN models by using knowledge distillation and boost the attacking success rate by 20\%. Gao et al.~\cite{gao2019boosting} integrated linear augmentation into substitute training and achieved success rates of 97.7\% and 92.8\% in MNIST and GTSRB classifiers. In addition, some researchers studied substitute attacks from a data perspective. Wang et al.~\cite{Wang2021delving} proposed a substitute training approach that designs the distribution of data used in the knowledge stealing process. Zhou et al.~\cite{Zhou2020dast} proposed a data-free substitute training method (DaST) to obtain substitute models by utilizing generative adversarial networks (GANs). However, the research on how to apply substitute-based adversarial attacks to the apps and the analysis of their performance is lacking. This paper fills the blank in this direction.

\subsection{Deep Learning Model Security of Apps }
Previous works on the security of deep learning models in mobile apps mainly focus on how to obtain information about the structure and weights of the model. Sun et al.~\cite{Sun2020MindYW} developed a pipeline that can analyze the model structure and weights and revealed that many deep learning models can not be extracted directly from mobile apps by decompiling the apks. Huang et al.~\cite{huang2021robustness} developed a pipeline to find the most similar model on TensorFlow Hub and then use it to generate adversarial examples. However, they still need to know the information of layers in the model to calculate similarity, which is used to find the most similar model on TensorFlow Hub for launching attacks. Different from Huang et al.'s work~\cite{huang2021robustness}, we do not require the knowledge of the structure and weights of the model. Li et al.~\cite{li2021deeppayload} proved the possibility to perform backdoor attacks on deep learning models and succeeded to use images with unique patterns to fool the models. To perform such attacks, Li et al.~\cite{li2021deeppayload} modified the internal structure of the model inside the app. Whereas, as a black-box approach, we do not need to alter anything inside the deep learning model. Different from existing works, our work focuses on investigating the security of the deep learning models in mobile apps in a complete black-box manner. Our work offers a new perspective and shed the light on security research in deep learning models for mobile.



\subsection{Adversarial Attacks and Defenses to Deep Learning Model}
Adversarial attacks show their power in spoofing deep learning models related to computer vision. Researches on image classification attack methods account for a large part. The most popular technique is adversarial image perturbations (AIP)~\cite{oh2017adversarial}. Dai et al.~\cite{dai2018adversarial} developed an attack method based on genetic algorithms and gradient descent, which demonstrates that the Graph Neural Network models are vulnerable to these attacks. Stepan Komkov and Aleksandr Petiushko~\cite{komkov2019advhat} proposed a reproducible technique to attack a real-world Face ID system. Baluja et al.~\cite{baluja2018learning} developed an Adversarial Transformation Network (ATN) to generate adversarial examples. It reaches a high success rate on MNIST-digit classifiers and Google ImageNet classifiers. 

As adversarial attacks pose a huge threat to deep learning models, researches on how to protect models from such attacks have also drawn wide attention. Li et al.~\cite{li2020defense} presented the gradient leaking hypothesis to motivate effective defense strategies. Liao et al.~\cite{liao2018defense} proposed a high-level representation guided denoiser (HGD) as a defense for deep learning models related to image classification. Zhou et al.~\cite{zhou2020adversarial} proposed a defense method to improve the robustness of DNN-based image ranking systems. Ciss{\'{e}} et al~\cite{moustapha2017parseval} proposed Parseval networks to improve robustness to adversarial examples. Alexandre et al.~\cite{araujo2020robust} used a Randomized Adversarial Training method to improve the robustness of deep learning neural networks.

However, even though there are a lot of researches on the adversarial attack and defense methods of deep learning models, research on this topic about mobile apps is scant. Our work proves that a black-box attack on the deep learning model inside the apps is feasible and provides a new perspective to evaluate the robustness of deep learning apps.

\section{Conclusion}
In this paper, we propose a practical black-box attack approach on deep learning apps and develop a corresponding pipeline. The experiment on 10 apps shows that the average attack success rate reaches 66.60\%. Compared with existing adversarial attacks on deep learning apps, our approach outperforms counterparts by 27.63\%. We also discuss how student model structure, teaching dataset size, and hyper-parameters of attack algorithms can affect attack performance.



\bibliographystyle{IEEEtran}

\bibliography{reference.bib}

\begin{thebibliography}{10}
\providecommand{\url}[1]{#1}
\csname url@samestyle\endcsname
\providecommand{\newblock}{\relax}
\providecommand{\bibinfo}[2]{#2}
\providecommand{\BIBentrySTDinterwordspacing}{\spaceskip=0pt\relax}
\providecommand{\BIBentryALTinterwordstretchfactor}{4}
\providecommand{\BIBentryALTinterwordspacing}{\spaceskip=\fontdimen2\font plus
\BIBentryALTinterwordstretchfactor\fontdimen3\font minus
  \fontdimen4\font\relax}
\providecommand{\BIBforeignlanguage}[2]{{%
\expandafter\ifx\csname l@#1\endcsname\relax
\typeout{** WARNING: IEEEtran.bst: No hyphenation pattern has been}%
\typeout{** loaded for the language `#1'. Using the pattern for}%
\typeout{** the default language instead.}%
\else
\language=\csname l@#1\endcsname
\fi
#2}}
\providecommand{\BIBdecl}{\relax}
\BIBdecl

\bibitem{xu2019first}
M.~Xu, J.~Liu, Y.~Liu, F.~X. Lin, Y.~Liu, and X.~Liu, ``A first look at deep
  learning apps on smartphones,'' in \emph{Proceedings of WWW}, 2019, pp.
  2125--2136.

\bibitem{tensorflowlite}
``Tensorflow lite,'' \url{https://tensorflow.google.cn/lite/}.

\bibitem{pytorchmobile}
``Pytorch mobile,'' \url{https://pytorch.org/mobile/home/}.

\bibitem{caff2mobile}
``Caffe2 mobile,'' \url{https://caffe2.ai/docs/mobile-integration.html}.

\bibitem{mindsporelite}
``Mindspore lite,'' \url{https://www.mindspore.cn/lite/en}.

\bibitem{coreml}
``Coreml,'' \url{https://developer.apple.com/machine-learning/core-ml/}.

\bibitem{firebaseOndeviceVSOncloud}
``Firebase-cloud vs on-device,''
  \url{https://firebase.google.com/docs/ml#cloud_vs_on-device}.

\bibitem{mcintosh2019can}
A.~McIntosh, S.~Hassan, and A.~Hindle, ``What can android mobile app developers
  do about the energy consumption of machine learning?'' \emph{Empirical
  Software Engineering}, vol.~24, no.~2, pp. 562--601, 2019.

\bibitem{kumar2020adversary}
C.~Kumar, R.~Ryan, and M.~Shao, ``Adversary for social good: Protecting
  familial privacy through joint adversarial attacks,'' in \emph{Proceedings of
  AAAI}, 2020, pp. 11\,304--11\,311.

\bibitem{Sun2020MindYW}
Z.~Sun, R.~Sun, L.~Lu, and A.~Mislove, ``Mind your weight(s): A large-scale
  study on insufficient machine learning model protection in mobile apps,'' in
  \emph{Proceedings of the 30th USENIX Security Symposium}, ser. Processings of
  USENIX Security, August 2021, pp. 1--17.

\bibitem{li2021deeppayload}
Y.~Li, J.~Hua, H.~Wang, C.~Chen, and Y.~Liu, ``Deeppayload: Black-box backdoor
  attack on deep learning models through neural payload injection,'' in
  \emph{Proceedings of ICSE-SEIP}, 2021, pp. 1--12.

\bibitem{huang2021robustness}
Y.~Huang, H.~Hu, and C.~Chen, ``Robustness of on-device models: Adversarial
  attack to deep learning models on android apps,'' in \emph{Proceedings of
  ICSE-SEIP}, 2021, pp. 1--12.

\bibitem{onlineartefacts}
``Online artefact,''
  \url{https://sites.google.com/view/blackbox-attack-on-dl-apps/home}.

\bibitem{yin1029feature}
X.~Yin, X.~Yu, K.~Sohn, X.~Liu, and M.~Chandraker, ``Feature transfer learning
  for face recognition with under-represented data,'' in \emph{Proceedings of
  CVPR}, 2019, pp. 5704--5713.

\bibitem{daniel2010towards}
D.~Zoran, M.~Chrzanowski, P.~Huang, S.~Gowal, A.~Mott, and P.~Kohli, ``Towards
  robust image classification using sequential attention models,'' in
  \emph{Proceedings of CVPR}, 2020, pp. 9480--9489.

\bibitem{chen2020unblind}
J.~Chen, C.~Chen, Z.~Xing, X.~Xu, L.~Zhut, G.~Li, and J.~Wang, ``Unblind your
  apps: Predicting natural-language labels for mobile gui components by deep
  learning,'' in \emph{Proceedings of ICSE}, 2020, pp. 322--334.

\bibitem{dong2020rtmobile}
P.~Dong, S.~Wang, W.~Niu, C.~Zhang, S.~Lin, Z.~Li, Y.~Gong, B.~Ren, X.~Lin, and
  D.~Tao, ``Rtmobile: Beyond real-time mobile acceleration of rnns for speech
  recognition,'' in \emph{Proceedings of DAC}, 2020, pp. 1--6.

\bibitem{wang2020next}
Q.~Wang, H.~Yin, T.~Chen, Z.~Huang, H.~Wang, Y.~Zhao, and N.~Q.~V. Hung, ``Next
  point-of-interest recommendation on resource-constrained mobile devices,'' in
  \emph{Proceedings of WWW}, 2020, pp. 906--916.

\bibitem{tensorflowhub}
``Tensorflow hub,'' \url{https://www.tensorflow.org/hub}.

\bibitem{mindspore}
``Mindspore,'' \url{https://www.mindspore.cn/lite/models/en}.

\bibitem{tan2020fastva}
T.~Tan and G.~Cao, ``Fastva: Deep learning video analytics through edge
  processing and npu in mobile,'' in \emph{Proceedings of INFOCOM}, 2020, pp.
  1947--1956.

\bibitem{goodfellow2015explaining}
I.~J. Goodfellow, J.~Shlens, and C.~Szegedy, ``Explaining and harnessing
  adversarial examples,'' in \emph{Proceedings of ICLR}, 2015, pp. 1--11.

\bibitem{Soot}
``Soot:a java optimization framework,'' \url{https://github.com/soot-oss/soot}.

\bibitem{flowdroid}
``Flowdroid,'' \url{https://github.com/secure-software-engineering/FlowDroid}.

\bibitem{tensorflowliteinference}
``Tensorflow lite inference,''
  \url{https://tensorflow.google.cn/lite/guide/inference#load_and_run_a_model_in_java}.

\bibitem{pytorchinference}
``Pytorch mobile inference,''
  \url{https://pytorch.org/mobile/android/#api-docs}.

\bibitem{DBLP:conf/gcce/TakedaYM20}
H.~Takeda, S.~Yoshida, and M.~Muneyasu, ``Learning from noisy labeled data
  using symmetric cross-entropy loss for image classification,'' in
  \emph{Proceedings on GCCE}, 2020, pp. 709--711.

\bibitem{simonyan2014very}
K.~Simonyan and A.~Zisserman, ``Very deep convolutional networks for
  large-scale image recognition,'' in \emph{Proceedings of ICLR}, 2015, pp.
  1--14.

\bibitem{liu2017delving}
Y.~Liu, X.~Chen, C.~Liu, and D.~Song, ``Delving into transferable adversarial
  examples and black-box attacks,'' in \emph{Proceedings of ICLR}, 2017, pp.
  1--14.

\bibitem{tramer2020ensemble}
F.~Tram{\`{e}}r, A.~Kurakin, N.~Papernot, I.~J. Goodfellow, D.~Boneh, and P.~D.
  McDaniel, ``Ensemble adversarial training: Attacks and defenses,'' in
  \emph{Proceedings of ICLR}, 2018, pp. 1--22.

\bibitem{kim2020torchattacks}
H.~Kim, ``Torchattacks: A pytorch repository for adversarial attacks,''
  \emph{arXiv preprint arXiv:2010.01950}, pp. 1--6, 2020.

\bibitem{dong2018boosting}
Y.~Dong, F.~Liao, T.~Pang, H.~Su, J.~Zhu, X.~Hu, and J.~Li, ``Boosting
  adversarial attacks with momentum,'' in \emph{Proceedings of CVPR}, 2018, pp.
  9185--9193.

\bibitem{madry2017towards}
A.~Madry, A.~Makelov, L.~Schmidt, D.~Tsipras, and A.~Vladu, ``Towards deep
  learning models resistant to adversarial attacks,'' in \emph{Proceedings of
  ICLR}, 2018, pp. 1--28.

\bibitem{zhang2019theoretically}
H.~Zhang, Y.~Yu, J.~Jiao, E.~P. Xing, L.~E. Ghaoui, and M.~I. Jordan,
  ``Theoretically principled trade-off between robustness and accuracy,'' in
  \emph{Proceedings of ICML}, 2019, pp. 7472--7482.

\bibitem{kurakin2016adversarial}
A.~Kurakin, I.~J. Goodfellow, and S.~Bengio, ``Adversarial examples in the
  physical world,'' in \emph{Proceedings of ICLR}, 2017, pp. 1--15.

\bibitem{imagenet}
``Imagenet,'' \url{http://www.image-net.org/challenges/LSVRC/2012/index}.

\bibitem{torchsgd}
``Pytorch sgd,''
  \url{https://pytorch.org/docs/master/generated/torch.optim.SGD.html}.

\bibitem{torchcross}
``Pytorch crossentropyloss,''
  \url{https://pytorch.org/docs/stable/generated/torch.nn.CrossEntropyLoss.html}.

\bibitem{kaggledatasetfruit}
``Kaggle dataset-fruit,''
  \url{https://www.kaggle.com/sriramr/fruits-fresh-and-rotten-for-classification}.

\bibitem{kai2016deep}
K.~He, X.~Zhang, S.~Ren, and J.~Sun, ``Deep residual learning for image
  recognition,'' in \emph{Proceedings of CVPR}, 2016, pp. 770--778.

\bibitem{huang2017densely}
G.~Huang, Z.~Liu, L.~Van Der~Maaten, and K.~Q. Weinberger, ``Densely connected
  convolutional networks,'' in \emph{Proceedings of CVPR}, 2017, pp.
  4700--4708.

\bibitem{mark2018mobilenet}
M.~Sandler, A.~G. Howard, M.~Zhu, A.~Zhmoginov, and L.~Chen, ``Mobilenetv2:
  Inverted residuals and linear bottlenecks,'' in \emph{Proceedings of CVPR},
  2018, pp. 4510--4520.

\bibitem{ma2018shufflenet}
N.~Ma, X.~Zhang, H.-T. Zheng, and J.~Sun, ``Shufflenet v2: Practical guidelines
  for efficient cnn architecture design,'' in \emph{Proceedings of ECCV}, 2018,
  pp. 116--131.

\bibitem{iandola2016squeezenet}
F.~N. Iandola, S.~Han, M.~W. Moskewicz, K.~Ashraf, W.~J. Dally, and K.~Keutzer,
  ``Squeezenet: Alexnet-level accuracy with 50x fewer parameters and< 0.5 mb
  model size,'' \emph{arXiv preprint arXiv:1602.07360}, pp. 1--13, 2016.

\bibitem{szegedy2015going}
C.~Szegedy, W.~Liu, Y.~Jia, P.~Sermanet, S.~Reed, D.~Anguelov, D.~Erhan,
  V.~Vanhoucke, and A.~Rabinovich, ``Going deeper with convolutions,'' in
  \emph{Proceedings of CVPR}, 2015, pp. 1--9.

\bibitem{tfdataset}
``Tensorflow dataset,''
  \url{https://www.tensorflow.org/datasets/catalog/tf_flowers}.

\bibitem{kaggledatasetroad}
``Kaggle dataset-road,''
  \url{https://www.kaggle.com/virenbr11/pothole-and-plain-rode-images}.

\bibitem{kaggledatasetpokemon}
``Kaggle dataset-pokemon,''
  \url{https://www.kaggle.com/lantian773030/pokemonclassification}.

\bibitem{hamdataset}
``Ham10000,''
  \url{https://dataverse.harvard.edu/dataset.xhtml?persistentId=doi:10.7910/DVN/DBW86T}.

\bibitem{googleimages}
``Google images,'' \url{https://images.google.com/}.

\bibitem{dong2021sentence}
J.~Dong, Z.~Guan, L.~Wu, X.~Du, and M.~Guizani, ``A sentence-level text
  adversarial attack algorithm against iiot based smart grid,'' \emph{Computer
  Networks}, pp. 1--11, 2021.

\bibitem{wang2019natural}
X.~Wang, H.~Jin, and K.~He, ``Natural language adversarial attacks and defenses
  in word level,'' \emph{arXiv preprint arXiv:1909.06723}, pp. 1--16, 2019.

\bibitem{dong2019efficient}
Y.~Dong, H.~Su, B.~Wu, Z.~Li, W.~Liu, T.~Zhang, and J.~Zhu, ``Efficient
  decision-based black-box adversarial attacks on face recognition,'' in
  \emph{Proceedings of CVPR}, 2019, pp. 7714--7722.

\bibitem{sun2018survey}
L.~Sun, M.~Tan, and Z.~Zhou, ``A survey of practical adversarial example
  attacks,'' \emph{Cybersecur.}, pp. 1--9, 2018.

\bibitem{Boloor2019simple}
A.~Boloor, X.~He, C.~D. Gill, Y.~Vorobeychik, and X.~Zhang, ``Simple physical
  adversarial examples against end-to-end autonomous driving models,'' in
  \emph{Proceedings of ICESS}, 2019, pp. 1--7.

\bibitem{Xu2020what}
X.~Xu, J.~Chen, J.~Xiao, L.~Gao, F.~Shen, and H.~T. Shen, ``What machines see
  is not what they get: Fooling scene text recognition models with adversarial
  text images,'' in \emph{Proceedings of CVPR}, 2020, pp. 12\,304--12\,314.

\bibitem{cui2020substitute}
W.~Cui, X.~Li, J.~Huang, W.~Wang, S.~Wang, and J.~Chen, ``Substitute model
  generation for black-box adversarial attack based on knowledge
  distillation,'' in \emph{Proceedings of ICIP}, 2020, pp. 648--652.

\bibitem{gao2019boosting}
X.~Gao, Y.-a. Tan, H.~Jiang, Q.~Zhang, and X.~Kuang, ``Boosting targeted
  black-box attacks via ensemble substitute training and linear augmentation,''
  \emph{Applied Sciences}, pp. 1--14, 2019.

\bibitem{Wang2021delving}
W.~Wang, B.~Yin, T.~Yao, L.~Zhang, Y.~Fu, S.~Ding, J.~Li, F.~Huang, and X.~Xue,
  ``Delving into data: Effectively substitute training for black-box attack,''
  in \emph{Proceedings of the IEEE/CVF Conference on Computer Vision and
  Pattern Recognition (CVPR)}, 2021, pp. 4761--4770.

\bibitem{Zhou2020dast}
M.~Zhou, J.~Wu, Y.~Liu, S.~Liu, and C.~Zhu, ``Dast: Data-free substitute
  training for adversarial attacks,'' in \emph{Proceedings of the IEEE/CVF
  Conference on Computer Vision and Pattern Recognition (CVPR)}, 2020, pp.
  234--243.

\bibitem{oh2017adversarial}
S.~J. Oh, M.~Fritz, and B.~Schiele, ``Adversarial image perturbation for
  privacy protection a game theory perspective,'' in \emph{Proceedings of
  ICCV}, 2017, pp. 1491--1500.

\bibitem{dai2018adversarial}
H.~Dai, H.~Li, T.~Tian, X.~Huang, L.~Wang, J.~Zhu, and L.~Song, ``Adversarial
  attack on graph structured data,'' in \emph{Proceedings of ICML}, 2018, pp.
  1115--1124.

\bibitem{komkov2019advhat}
S.~Komkov and A.~Petiushko, ``Advhat: Real-world adversarial attack on arcface
  face {ID} system,'' \emph{CoRR}, pp. 1--9, 2019.

\bibitem{baluja2018learning}
S.~Baluja and I.~Fischer, ``Learning to attack: Adversarial transformation
  networks,'' in \emph{Proceedings of AAAI}, 2018, pp. 1--13.

\bibitem{li2020defense}
Y.~Li, S.~Cheng, H.~Su, and J.~Zhu, ``Defense against adversarial attacks via
  controlling gradient leaking on embedded manifolds,'' in \emph{Proceedings of
  ECCV}, 2020, pp. 753--769.

\bibitem{liao2018defense}
F.~Liao, M.~Liang, Y.~Dong, T.~Pang, X.~Hu, and J.~Zhu, ``Defense against
  adversarial attacks using high-level representation guided denoiser,'' in
  \emph{Proceedings of CVPR}, 2018, pp. 1778--1787.

\bibitem{zhou2020adversarial}
M.~Zhou, Z.~Niu, L.~Wang, Q.~Zhang, and G.~Hua, ``Adversarial ranking attack
  and defense,'' in \emph{Proceedings of ECCV}, 2020, pp. 781--799.

\bibitem{moustapha2017parseval}
M.~Ciss{\'{e}}, P.~Bojanowski, E.~Grave, Y.~N. Dauphin, and N.~Usunier,
  ``Parseval networks: Improving robustness to adversarial examples,'' in
  \emph{Proceedings of ICML}, 2017, pp. 854--863.

\bibitem{araujo2020robust}
A.~Araujo, R.~Pinot, B.~N{\'{e}}grevergne, L.~Meunier, Y.~Chevaleyre, F.~Yger,
  and J.~Atif, ``Robust neural networks using randomized adversarial
  training,'' \emph{CoRR}, pp. 1--9, 2019.

\end{thebibliography}

\end{document}